\documentclass[a4paper,11pt]{article}
\pdfoutput=1

\usepackage{jheppub} 
\usepackage[T1]{fontenc} 

\usepackage[utf8]{inputenc}
\usepackage{todonotes}
\usepackage{amssymb,amsmath,amsbsy}
\usepackage{braket}
\usepackage{graphicx,color}

\newcommand{\ben}{\begin{enumerate}}
\newcommand{\een}{\end{enumerate}}

\newcommand{\be}{\begin{equation}}
\newcommand{\ee}{\end{equation}}
\newcommand{\bea}{\begin{eqnarray}}
\newcommand{\eea}{\end{eqnarray}}
\newcommand{\lb}{\left(}
\newcommand{\rb}{\right)}

\newcommand{\mA}{\mathcal{A}}

\newcommand{\nn}{\nonumber}
 
\newcommand{\mH}{\mathcal{H}}

\newcommand{\nbox}{{\,\lower0.9pt\vbox{\hrule \hbox{\vrule height 0.2 cm \hskip 0.19 cm \vrule height 0.2 cm}\hrule}\,}}

\title{\boldmath Modular zero modes and sewing the states of QFT}

\author[a,b]{Nima Lashkari}
 \affiliation[a]{Department of Physics and Astronomy, Purdue University, West Lafayette, IN
47907, USA}
\affiliation[b]{School of Natural Sciences, Institute for Advanced Study, Princeton, New Jersey 08540, USA}
 
 \emailAdd{nima@purdue.edu}

\abstract{We point out an important difference between continuum relativistic quantum field theory (QFT) and lattice models with dramatic consequences for the theory of multi-partite entanglement. On a lattice given a collection of density matrices $\rho^{(1)},\rho^{(2)}, \cdots, \rho^{(n)}$ there is no guarantee that there exists an $n$-partite pure state $\ket{\Omega}_{12\cdots n}$ that reduces to these marginals. The state $\ket{\Omega}_{12\cdots n}$ exists only if the eigenvalues of the density matrices $\rho^{(i)}$ satisfy certain polygon inequalities. We show that in QFT, as opposed to lattice systems, splitting the space into $n$ non-overlapping regions any collection of local states $\omega^{(1)},\omega^{(2)},\cdots \omega^{(n)}$ come from the restriction of a global pure state. The reason is that rotating any local state $\omega^{(i)}$ by unitary $U_i$ localized in the $i^{th}$ region we come arbitrarily close to any other local state $\psi^{(i)}$. We construct explicit examples of such local unitaries using the cocycle.}

\begin{document}

\maketitle

\flushbottom

\section{Introduction}
Consider the Hilbert space of an $n$-partite lattice quantum system $ \mathcal{H}_1\otimes\mathcal{H}_2\otimes \cdots \mathcal{H}_n$, where $\mathcal{H}_i$ is the Hilbert space of site $A_i$. We call a pure state of the $n$-partite system a global state, and the density matrices $\rho^{(i)}$ at site $A_i$ local states.
Given a collection of local states $\rho^{(1)}$ to $\rho^{(n)}$ we say we can {\it sew} them if there exists a global pure state  $\ket{\Omega}$ that reduces to $\rho^{(i)}$ at each site $A_i$; e.g. $\rho^{(i)}=\text{tr}_{j\neq i}\ket{\Omega}\bra{\Omega}$ for all $i$.
Physically, the collection of density matrices on $A_i$ are a mean field approximation to a pure global state.
In finite quantum systems, there are constraints on the density matrices that can be sewn together. For instance, in a bipartite system, the density matrices $\rho^{(1)}$ and $\rho^{(2)}$ can be sewn if and only if they have the same eigenvalues. These eigenvalues are are the Schmdit coefficients of the pure global state $\ket{\Omega}\in\mH_1\otimes\mH_2$. 
The constraints become more complicated for $n>2$, and imply that the eigenvalues of $\rho^{(i)}$ should satisfy polygon inequalities \cite{bravyi2003requirements,higuchi2003one}.
The knowledge of $\rho^{(j)}$ for $j\neq i$ and the fact that the global state is pure implies some information about $\rho^{(i)}$, however this information is partial because it does not fix $\rho^{(i)}$. A local unitary rotation $U_i$ at site $A_i$ does not affect the state on the other sites, therefore $\rho^{(i)}$ can always be replaced with $U_i \rho^{(i)} U_i^\dagger$.\footnote{If the local Hilbert space $\mH_i$ is infinite dimensional it suffices for $U_i$ to be an isometry, i.e. $U_i^\dagger U_i=1_i$ and $U_iU_i^\dagger$ a projection in $A_i$. In finite dimensions, there are no distinctions between unitaries and isometries. In infinite dimensional systems every local isometry  has a unitary arbitrarily close to it in norm topology. We prove this in appendix \ref{AppC}. We thank Martin Argerami and Roberto Longo for discussion on this.} 

Consider a pure state of a Poincare invariant QFT in Minkowski spacetime of arbitrary dimensions on a constant time-slice. Split the time-slice into $n$ non-overlapping regions $A_1$ to $A_n$ such that $\cup_i A_i$ is the whole time-slice. Each region $A_i$ has a local algebra of observables $\mathcal{A}_i$. The analog of local density matrices are the restrictions of the the global state to the local algebras:
\begin{eqnarray}\label{rholocal}
\omega_i(a_i)=\braket{\Omega|a_i\Omega}\qquad \forall a_i\in\mA_i\ .
\end{eqnarray} 
Since the global algebra of QFT is not the tensor product $\mA_1\otimes\cdots \mA_n$ we cannot sew all states perfectly. However, in this work we argue that as opposed to lattice systems, if we allow for arbitrarily small errors it is possible to sew any collection of $\omega_i$s in a pure state. As we show, the reason is that, in QFT, for any local state $\psi_i$ there exists a unitary $U_i$ in $\mathcal{A}_i$ such that $\psi_i(a_i)\simeq \omega_i(U_i^\dagger a_i U_i)$ for all $a_i\in\mathcal{A}_i$. We use $\simeq$ as opposed to equality to indicate that the unitarily rotated states approximate $\omega_i$ arbitrarily well in the norm topology of operators. For instance, this procedure allows us to start with an arbitrary global state $\ket{\Psi}$ and act on $A$ with a unitary to prepare a state that is the same as vacuum on region $A$ and the same as $\ket{\Psi}$ outside of $A$. Similarly, one can sew states of different QFTs together. The conventional formulation of quantum entanglement theory is designed for lattice systems. To study entanglement in QFT, one needs to replace the conventional formulation with a more general algebraic framework called Modular theory that applies to any quantum system from qubits to QFT. We use modular theory to construct the local unitaries that sew states to the vacuum in QFT. This unitary was recently used to derive a surprising inequality that holds in any QFT called Quantum Null Energy Condition (QNEC), but has no analog in lattice systems \cite{Ceyhan:2018zfg}.

\section{Sewing qudits}

It is worthwhile to analyze the sewing problem on a lattice first to identity the origin of the obstruction to sew arbitrary states. We focus on the bipartite case, and come back to the general $n$-partite case in section \ref{sewmulti}. 
Every density matrix of system $A$ written in its eigenbasis $\rho=\sum_k p_k\ket{k}\bra{k}$ has a canonical purification in a pure global state $\ket{\Omega}=\sum_k \sqrt{p}_k\ket{k,k}\in \mH_A\otimes \mH_{\bar{A}}$, where $\mH_{\bar{A}}$ is isomorphic to $\mH_A$ and the reduced state on $\bar{A}$ is also $\omega$. The probability $p_k$ corresponds to the projections $e_k=\ket{k}\bra{k}$ and if all $p_k>0$ we say the state has the Reeh-Schlieder property. Reeh-Schlieder states have invertible density matrices and form a dense set of states. In the remainder of this work, we restrict to the Reeh-Schlieder states. The generalization of our discussion to non-Reeh-Schlieder states is straightforward.
The canonical purification of a density matrix $\rho$ sews it to a second identical copy on $\bar{A}$. This symmetry of the canonical purification $\ket{\Omega}$ is captured by its invariance under the action of the anti-linear operator $J_\Omega$ that swaps the eigenbasis of $\rho_A$ and $\rho_{\bar{A}}$: $J_\Omega(c\ket{k,k'})=c^*\ket{k',k}$. This operator is called the modular conjugation and depends on the state $\ket{\Omega}$ through the choice of the eigenbasis of $\rho$. Given a second density matrix $\psi=\sum_k q_k\ket{\alpha_k}\bra{\alpha_k}$ there is a unique purification $\ket{\Psi_\Omega}=(\psi^{1/2}\otimes \rho^{-1/2})\ket{\Omega}$ that is symmetric under the swap $J_\Omega\ket{\Psi_\Omega}=\ket{\Psi_\Omega}$. 
We call $\Delta_{\Psi|\Omega}\equiv \psi\otimes \rho^{-1}$ the relative modular operator and write $\ket{\Psi_\Omega}=\Delta_{\Psi|\Omega}^{1/2}\ket{\Omega}$. If both density matrices are the same we write $\Delta_\rho=\rho\otimes\rho^{-1}$ and call it the modular operator.
These operators act in the global Hilbert space of $A\bar{A}$. For two vectors $\ket{\Psi_\Omega}$ and $\ket{\Phi_\Omega}$ both invariant under $J_\Omega$ we have
\begin{eqnarray}\label{naturalcone}
\Delta_{\Psi_\Omega|\Phi_\Omega}^{1/2}\ket{\Phi_\Omega}=\ket{\Psi_\Omega}\ .
\end{eqnarray}
The adjoint action of the modular conjugation on operators in $A$ sends them to $\bar{A}$ and vice versa:
\begin{eqnarray}
J_\Omega(\ket{k}\bra{k'}_A\otimes 1_{\bar{A}})J_\Omega=1_A\otimes \ket{k}\bra{k'}_{\bar{A}}\ .
\end{eqnarray}
If $a$ is in the algebra of $A$ then we define $a_J=J_\Omega a J_\Omega$ to be its corresponding operator in $\bar{A}$. Hereafter, we suppress the subscript $\Omega$ of $J$, and where it is clear from the context suppress the identity operator $1_{\bar{A}}$ to simplify notation.

In the bipartite case, the necessary and sufficient condition for sewing two density matrices is that they have the same eigenvalues. If $\psi$ does not have the same eigenvalue as $\rho$ there might still exist a density matrix very close to $\psi$ that can be sewn to $\rho$.
We would like to find the closest density matrix to $\psi$ that can be sewn to $\rho$. The only density matrices that can be sewn to $\rho$ are $U\rho U^\dagger$, therefore we have to introduce a distance measure $d(\rho,\psi)$ on the space of density matrices and taking an infimum over all unitaries: $\inf_{U}d(U\rho U^\dagger,\psi)$.
The trace distance is a distance measure on the space of density matrices defined to be
\begin{eqnarray}\label{tracedis}
\|\rho-\psi\|_1^2=\text{tr}\lb |\rho-\psi|\rb=\sup_{\|C\|_{\infty}=1}\text{tr}\lb C (\rho-\psi)\rb
\end{eqnarray}
where $\|C\|_\infty$ is the operator norm of $C$ that is its largest eigenvalue. 
It is sometimes more convenient to work with the following distance measure on the space of density matrices:
\begin{eqnarray}\label{disFro}
&&d_F(\rho,\psi)^2=\frac{1}{2}\|\sqrt{\rho}-\sqrt{\psi}\|_F^2=1-\text{tr}\lb \sqrt{\rho}\sqrt{\psi}\rb
\end{eqnarray}
where $\|X\|_F^2=\text{tr}\lb X^\dagger X\rb$ is the Frobenius norm of a matrix. If two density matrices are close in the Frobenius distance they are also close in trace distance; see Appendix \ref{AppA}. The advantage of the Frobenius distance is that it matches the Hilbert space distance of the canonical purifications:
\begin{eqnarray}\label{Hilbertdis}
&&d_F(\rho,\psi)=\|\ket{\Psi_\Omega}-\ket{\Omega}\|\ .
\end{eqnarray}
If we think of $\ket{\chi}=\ket{\Psi_\Omega}-\ket{\Omega}$ as an unnormalized vector in the Hilbert space we have 
\begin{eqnarray}\label{norm}
d_F(\rho,\psi)=\|\ket{\chi}\|=\sup_{\|\ket{\Phi}\|=1}|\braket{\Phi|\chi}|\ .
\end{eqnarray}
The vector $\ket{\Phi}$ that achieves the supremum is parallel to $\ket{\chi}$ and saturates the Cauchy-Schwarz inequality. On a lattice, the infimum distance $\inf_U d_F(U\rho U^\dagger, \psi)$ can be calculated explicitly.
The unitary $U_0=\sum_k \ket{\alpha_k}\bra{k}$ that rotates the eigenbasis of $\rho$ to that of $\psi$ sets an upper bound on this infimum distance:
\bea\label{inf}
\inf_U d(U\rho U^\dagger,\psi)\leq d(p,q)
\eea 
where $d(p,q)$ is the classical analog of our distance measure for the probability distributions $\{p_k\}$ and $\{q_k\}$, i.e. $d(p,q)^2=\frac{1}{2}\sum_k (\sqrt{q_k}-\sqrt{p_k})^2=1-\sum_k \sqrt{p_kq_k}$.
In general, this is not a tight bound because the unitaries that relabel the basis of $\rho$ can further reduce the distance. Each permutation in the symmetric group  $\sigma\in S_d$, where $d$ is the dimension of $\mH_A$, gives a relabelling unitary $U_\sigma=\sum_k\ket{\alpha_{\sigma(k)}}\bra{\alpha_k}$. The action of these unitaries is equivalent to keeping the eigenvectors fixed and permuting the eigenvalues. We tighten our bound to
\bea\label{inf2}
&&\inf_Ud(\rho,\psi)\leq \inf_{\sigma\in S_d}d(p,\sigma(q))\nn\\
&&\sigma(q)_k=q_{\sigma(k)}\ .
\eea
The classical distance $d(p,\sigma(q))$ is minimized if the fidelity $\sum_k\sqrt{p_kq_k}$ is maximized. If we order $p_k$ in decreasing order $p_1\geq p_2\geq\cdots \geq p_d$ this maximum is achieved by the relabeling unitary that orders $q$ in decreasing order: $q_1\geq q_2\geq \cdots \geq q_d$, and then matches them by $\sigma(k)=k$:
\bea\label{equalinf}
&&\inf_{\sigma \in S_d} d^2_F(p,\sigma(q))=1-\sum_k \sqrt{p_k q_k}=\sum_k q_k\lb 1-\frac{p_k}{q_k}\rb,
\eea
The terms $p_k/q_k$ above are the eigenvalues of the relative modular operator. The Frobenius distance is small if the spectrum of the relative modular operator is limited to a small range near one. 
In fact, for finite $d$ the inequality in (\ref{inf2}) is an equality \cite{zhang2014quantum}, and the minimum of the classical distance we found is the same as the minimum of the quantum distance. 

In infinite dimensions, the spectrum of the modular operator can become continuous and notion of eigenvalues might stop making sense. In such a case, there can exist relabelling unitaries that squeeze the spectrum of the relative modular operator to a small neighborhood of one.
These are the unitaries that we allow us to sew any two local states in QFT. We come back to this in section \ref{secinv}.

\section{Infinite dimensions}

The generalization of the trace distance in (\ref{tracedis}) to local states in QFT is 
\begin{eqnarray}\label{tracedisQFT}
\|\omega-\psi\|_1^2=\sup_{\|a\|=1}|\omega(a)-\psi(a)|
\end{eqnarray}
with $\|a\|$ is the operator norm and $\omega$ is the restriction of the global state to a local algebra as in (\ref{rholocal}). Let us call the set of states $U\omega U^\dagger$ for $U$ unitaries in $\mA$ the unitary orbit of $\omega$, and denote by $[\omega]$ the equivalence class of $\omega$ under unitary rotations.
We define the strong unitary orbit distance to be
\begin{eqnarray}
d_s([\omega],[\psi])=\inf_U\sup_{\|a\|=1}|\omega(a)-\psi(U^\dagger a U)|\ .
\end{eqnarray}
where $U$ are unitaries in the local algebra. 
As in the case of density matrices, it is more convenient to work the Frobenius distance in (\ref{Hilbertdis}) of the canonical purifications and use the inequality
\begin{eqnarray}
\|\ket{\Omega}-\ket{\Psi_\Omega}\|^2\leq \|\omega-\psi\|^2\leq 2\|\ket{\Omega}-\ket{\Psi_\Omega}\|
\end{eqnarray}
and consider the unitary orbit distance measure 
\begin{eqnarray}
d_F([\omega],[\psi])=\inf_U\|\ket{U_\Omega}-\ket{\Psi_\Omega}\|^2\ .
\end{eqnarray}
Both $\ket{U_\Omega}=UJ U\ket{\Omega}=UU_J\ket{\Omega}$ and $\ket{\Psi_\Omega}$ are the canonical purifications of local states $U^\dagger\omega U$ and $\psi$, respectively, and are both invariant under the action of $J$. We apply (\ref{naturalcone}) to rewrite the Frobenius distance in terms of the relative modular operator:
\begin{eqnarray}\label{weakstrong}
d_F([\omega], [\psi])&&=\inf_{U\in\mA}\left\|\lb\Delta_{\Psi_\Omega|U_\Omega}^{1/2}-1\rb \ket{U_\Omega}\right\|\nn\\
&&=\inf_{U\in\mA}\|(\Delta_{\Psi|\Omega}^{1/2}-1)U\ket{\Omega}\|\ .
\end{eqnarray}
The second line can be checked explicitly in the case of density matrices, or more formally using the equations $\ket{\Psi_\Omega}=V'\ket{\Psi}$ with $V'$ and $U'$ unitaries in the algebra of $\bar{A}$ and $\Delta_{V'\Psi|UU'\Omega}=U'\Delta_{\Psi|\Omega}U$ that can be shown using equation (\ref{relativeTomita}). The unitary orbit distances is zero if and only if the state $\ket{U_\Omega}$ is invariant under the action of the relative modular operator. This ties the problem of  
sewing local states $\psi$ of $A$ to $\omega$ of its complementary region $\bar{A}$ to the problem of locally preparing an invariant state of their relative modular operator. 

In infinite dimensions, the infimum and supremum above need not commute. Therefore, we define a second distance measure that we call the weak unitary orbit distance to be
\begin{eqnarray}\label{distanceweak}
d_w([\omega],[\psi])=\sup_{\|a\|=1}\inf_U|\omega(a)-\psi(U^\dagger a U)|
\end{eqnarray}

To obtain intuition about the difference between the weak and the strong distance consider the following example. Take the Hilbert space of a particle on a line and the set of unitaries $U_x=e^{i x P}$ with $P$ the translation operator and the matrix element $\braket{\Phi|U_x\Psi}$. For generic $\ket{\Phi}$ and $\ket{\Psi}$ we have
\begin{eqnarray}
&&\lim_{x\to\infty}\sup_{\ket{\Phi}\in\mH}|\braket{\Phi|e^{i x P}\Psi}|=\lim_{x\to\infty} \|e^{ix P}\ket{\Psi}\|=1
\end{eqnarray}
whereas if we commute the $\sup$ and $\lim$ we find
\begin{eqnarray}\label{limitweak}
&&\sup_{\ket{\Phi}\in\mH}\lim_{x\to\infty}|\braket{\Phi|e^{i x P}\Psi}=\braket{\Phi|\Pi_0\Psi}<1
\end{eqnarray}
where $\Pi_0$ is the projection to the subspace of translation-invariant states: $P\Pi_0=0$. Before proving this statement, let us explain why it holds in physics terms.
For vectors $\ket{\Phi}$ and $\ket{\Psi}$ that are the wave packets of two particles localized at some point in space with some width and the operator $e^{i x P}$ moves them away from each other. In the limit of infinite distance, the overlap between the wave-packets decays to zero. 
Now, if we replace $\ket{\Psi}$ with the delta-function normalizable momentum eigenstate $\ket{p}$ then $\braket{\Phi|e^{i x P}|p}=\braket{\Phi|p} e^{i x P}$ that oscillates erratically at large $t$ with no limit. The point is that to make physical states with unit norm we need wave-packets for which the limit in (\ref{limitweak}) exists. In other words, we have the limit in (\ref{limitweak}) because the translation operator has an entirely continuum spectrum with no normalizable eigenvectors. The momentum eigenstate $\ket{p}$ is not normalizable.

If we call $f(x)=|\braket{\Phi|e^{i x P}\Psi}|$ then the large time limit in (\ref{limitweak}) is the infinite Fourier mode of $f(x)$. The spectrum of the momentum operator is continuous, therefore
\begin{eqnarray}
&&\lim_{x\to\infty}|\braket{\Phi|e^{i x P}\Psi}=\int_{-\infty}^{\infty} e^{i x p} \braket{\Phi|\Pi_p(dp)\Psi}
\end{eqnarray}
where $\Pi_p(dp)$ is a projection operator valued measure. In the language of delta-function normalizable vectors $\Pi_p(dp)=dp \ket{p}\bra{p}$. Set $\ket{\Phi}=\ket{\Psi}$ so that $\hat{f}(p)$ is positive and
\begin{eqnarray}
\int |\hat{f}(p)| dp=\braket{\Psi|\Psi}=1\ .
\end{eqnarray}
Then, equation (\ref{limitweak}) for $\ket{\Phi}=\ket{\Psi}$ follows from the Riemann-Lebesgue lemma that says if $\int |\hat{f}(p)|dp<\infty$ for function $\hat{f}(p)$ its infinite Fourier mode $\lim_{x\to\infty}f(x)$ vanishes. The equation (\ref{limitweak}) for all matrix elements follows from considering the diagonal terms in states $\ket{\Phi}\pm \ket{\Psi}$. 

One can ask how large is the subspace of translation invariant states $\Pi_0\mH$. If there is an operator $h$ that commutes with the group $e^{it P}$ and $\ket{\Omega}$ is a translation invariant state then $h\ket{\Omega}$ is also translation invariant, and any translation invariant state can be written as $h\ket{\Omega}$ for some $h$ that commutes with the group $e^{it P}$.
If we consider the unitary flow $U_t=e^{i t K_x}$ where $K_x$ is boost the only invariant state is the vacuum, and $\Pi_0$ is the projection to the vacuum $\ket{\Omega}\bra{\Omega}$.

To come closer to the expression in (\ref{distanceweak}) consider the expectation value
$\lim_{x\to \infty}\braket{\Psi|a(x)\Psi}$
where $a(x)=e^{i x P}ae^{-i x P}$ is the translation of the operator $a$ and the excited state is $\ket{\Psi}=b\ket{\Omega}$. Then,
\begin{eqnarray}
&&\lim_{x\to \infty}\braket{\Psi|a(x)\Psi}=\lim_{x\to\infty}\braket{\Psi|a(x) b\Omega}\nn\\
&&=\lim_{x\to\infty}\braket{\Psi| b e^{i x P}a\Omega}=\braket{\Psi|b\Pi_0 a\Omega}
\end{eqnarray}
where we have used (\ref{limitweak}) the fact that the matrix elements of $[a(x),b]$ vanish in the limit $x\to\infty$ by causality. Consider the decompsoiton of the projection $\Pi_0$ into an orthonormal set of translation-invariant states  $\Pi_0=\sum_h \ket{h}\bra{h}$ with $\ket{h}=h\ket{\Omega}$. Then, the clustering property of correlation functions in state $\ket{\Omega}$ fails
\begin{eqnarray}
\lim_{x\to\infty}\braket{\Omega|a(x) b\Omega}\neq \braket{\Omega|a\Omega}\braket{\Omega| b\Omega}
\end{eqnarray}
unless $\ket{\Omega}$ is the unique translation-invariant state in which case
\begin{eqnarray}
\lim_{x\to\infty}\braket{\Psi|a(x)\Psi}=\braket{\Omega|a\Omega}
\end{eqnarray}
and the matrix elements of the operator $\lim_{a\to\infty}a(x)$ are the same as the identity operator multiplied by $\braket{\Omega|a\Omega}$. 

We are now ready to argue qualitatively why in QFT the weak unitary orbit distance vanishes. The rigorous proof is postponed to section \ref{modulartheory}. We use the intuition of density matrices and consider the unitary operators $\Delta_{\Omega}^{it}=\omega^{it}\otimes \omega^{-it}$, $\Delta_\Psi^{it}=\psi^{it}\otimes \psi^{-it}$ for density matrices $\omega$ and $\psi$. We define the unitary flow $u_{\omega|\psi}(t)=\omega^{it}\psi^{-it}\in \mA$ called the {\it cocycle} \cite{connes1973classification}. The claim is that the cocycle in the limit of $t\to\infty$ is the unitary that makes the weak unitary orbit distance defined in (\ref{distanceweak}) vanish. To simply the notation, we write $u_t=u_{\omega|\psi}(t)$ so that
\begin{eqnarray}
&&\psi(u_t^\dagger a u_t)=\braket{\Psi|(u_t^\dagger\otimes u_t) \:a\: (u_t\otimes u_t^\dagger)|\Psi}\nn\\
&&=\braket{\Psi|\Delta_\Omega^{-it} a\Delta_\Omega^{it}|\Psi}
\end{eqnarray}
where we have used 
\begin{eqnarray}
    (u_t\otimes u_t^\dagger)\ket{\Psi}&&=(\omega^{it}\otimes \omega^{-it})(\psi^{-it}\otimes \psi^{it}\ket{\Psi}\nn\\
    &&=(\omega^{it}\otimes \omega^{-it})\ket{\Psi}=\Delta_\Omega^{it}\ket{\Psi}\ .
\end{eqnarray}
The eigenvalues of the modular operator $\Delta_{\omega}=\sum_{kk'}\frac{p_k}{p_{k'}}\ket{kk'}\bra{kk'}$ are $p_k/p_{k'}$. The defining feature of the algebra of QFT is that the spectrum of the modular operator $\Delta_\Psi$ for any state is entirely continuous with no eigenvalues. Then, it is clear from (\ref{limitweak}) that if $\ket{\Omega}$ is the only invariant state of $\Delta_\Omega$ we have
\begin{eqnarray}
\lim_{t\to\infty}\psi(u_t^\dagger a u)=\lim_{t\to\infty}\braket{\Psi|\Delta_\Omega^{-it} a \Delta_\Omega^{it}\Psi}=\braket{\Omega|a\Omega}=\omega(a)\nn\ .
\end{eqnarray}
Therefore, $d_w([\omega],[\psi])=0$. In the vacuum of QFT, the modular operator of half-space is the boost, which has only one invariant state. 
Note that the cocycle cannot make the strong distance small because $(u_t\otimes u_t^\dagger)\ket{\Psi}\ket{\Psi}=\Delta_\Omega^{it}\ket{\Psi}$ and as a result
\begin{eqnarray}
\|(u_t\otimes u_t^\dagger)\ket{\Psi}-\ket{\Omega}\|^2&&=2\lb 1-\braket{\Omega|\Delta_\Omega^{it}\Psi}-\braket{\Psi|\Delta_\Omega^{-it}\Omega}\rb\nn\\
&&=\|\ket{\Psi}-\ket{\Omega}\|^2\ .
\end{eqnarray}
If the strong distance is small the weak distance is also small, whereas as we saw above the opposite is incorrect. The equation 
(\ref{weakstrong}) says that the unitary $U$ that sews local states $\omega$ and $\psi$ is the unitary that rotates the purification of $\omega$ to an invariant vector of the relative modular operator. In the next section, we study the invariant states of relative modular operator. 

\section{Invariant states of relative modular operator}\label{secinv}

To make the strong unitary orbit distance small one needs to find unitaries that prepare invariant states of the relative modular operator.
To find these invariant states we look at its spectrum of the relative modular operator in the case density matrices exist:
\begin{eqnarray}\label{specrel}
\Delta_{\Psi|\Omega}\equiv\sum_{\lambda}e^\lambda P_\lambda =\sum_{kk'}\frac{q_{k'}}{p_k}\ket{\alpha_{k'}k}\bra{\alpha_{k'}k},
\end{eqnarray}
where $e^\lambda$ are the eigenvalues $q_{k'}/p_k$ for any pair $(k,k')$ and $P_\lambda$ is the projection to the subspace of eigenvalue $e^\lambda$. There can be degeneracies for each $\lambda$: 
\begin{equation}
    P_\lambda=\sum_{\gamma=1}^{d_\lambda}\ket{\lambda;\gamma}\bra{\lambda;\gamma}\ .
\end{equation}
The invariant states of the relative modular operator are in the subspace $P_0\mH$. 
If there are no $q_{k'}=p_k$ the projection $P_0=0$, and there are no invariant states. If $S_0$ is the subset of pairs $(k,k')$ with $p_k=q_{k'}$ then
\begin{eqnarray}
P_0=\sum_{(k,k')\in S_0}\ket{\alpha_{k'},k}\bra{\alpha_{k'},k}\ .
\end{eqnarray}
The local operator $v=\sum_{(k,k')\in S_0}\ket{\alpha_{k'}}\bra{k}$ is a partial isometry\footnote{A partial isometry $v$ is an operator that satisfies $v^\dagger v=\pi_1$ and $vv^\dagger=\pi_2$ where $\pi_1$ and $\pi_2$ are projections. An operator $\pi$ is a projection if $\pi=\pi^\dagger$ and 
$\pi^2=\pi$.} of system $A$
that acts on $\ket{\Omega}$ and prepares invariant states of the relative modular operator
\begin{eqnarray}\label{partial}
    &&\Delta_{\Psi|\Omega}\lb v\otimes 1\rb\ket{\Omega}=\lb v\otimes 1\rb\ket{\Omega}\ .
\end{eqnarray}
The operator $v$ rotates the eigenvectors of $\omega$ to those of $\psi$ and further relabels them such that $q_{k'}=p_k$.
To simplify the discussion, let us order $p_k$ in descending order. 
The operator $v$ is unitary if and only if for every eigenvalue $p_k$ there exists a $k'$ such that $q_{k'}=p_k$, in which case, $v$ is the unitary that sorts $q_{k'}$ in descending order. We recover the statement that one can sew two density matrices $\omega$ and $\psi$ if and only if they have the same eigenvalues. As we saw in the last section, when the eigenvalues do not match, the unitary $v$ that sorts $q_{k'}$ still minimizes the unitary orbit distance.

In infinite systems, the eigenvalues $p_{k'}/p_k$ of the modular operator can become unbounded  for small enough $p_k$. 
Let us assume that our states of interest, $\ket{\Psi}$ and $\ket{\Omega}$, both have the property that the spectra of their modular operators are the entire continuum $[0,\infty)$ and the number of degeneracies at each $\lambda$ goes to infinity:
\begin{eqnarray}\label{modularop}
\Delta_\Omega&&=\sum_{kk'}p_kp_{k'}^{-1}\ket{kk'}\bra{kk'}=\int d\lambda_1\: e^{\lambda_1}\sum_\gamma \ket{\lambda_1;\gamma}\bra{\lambda_1,\gamma}\nn\\
\Delta_\Psi&&=\sum_{kk'}q_kq_{k'}^{-1}\ket{\alpha_k\alpha_{k'}}\bra{\alpha_k\alpha_{k'}}\nn\\
&&=\int d\lambda_2\: e^{\lambda_2}\sum_\gamma \ket{\lambda_2;\gamma}\bra{\lambda_2,\gamma},
\end{eqnarray}
keeping in mind that the vectors $\ket{\lambda;\gamma}$ are generalized eigenvectors that cannot be normalized:
$\braket{\lambda;\gamma|\lambda';\gamma'}=\delta_{\gamma\gamma'}\delta(\lambda-\lambda')$. 
In analogy with (\ref{partial}), we act on $\ket{\Omega}$ with the partial isometry $v_{k'k}=\ket{\alpha_{k'}}\bra{k}$ and prepare an eigenstate of the relative modular operator 
\begin{equation}
    \Delta_{\Psi|\Omega}\lb v_{k'k}\ket{\Omega}\rb=q_{k'}p_k^{-1}\lb v_{k'k}\ket{\Omega}\rb\ .
\end{equation}
The partial isometry $f_{lk'}=\ket{\alpha_l}\bra{\alpha_{k'}}$ that relabels $q_{k'}$ to $q_l$ acts on $v_{k'k}\ket{\Omega}$ and creates a new eigenstate
\begin{eqnarray}
\Delta_{\Psi|\Omega}\lb f_{lk'} v_{k'k}\ket{\Omega}\rb=q_{l}p_k^{-1}\lb f_{lk'} v_{k'k}\ket{\Omega}\rb\ .
\end{eqnarray}
This partial isometry sends an eigenspaces of relative modular operator to another one simply by changing the label:
\begin{eqnarray}
f_{lk}P_\lambda\mathcal{H}\in P_{\lambda q_l/q_{k}}\mathcal{H}\ .
\end{eqnarray}
Since we assumed that the spectrum of $\Delta_\Psi$ is continuous and entire in $[0,\infty)$ we can tune the partial isometry $f$ such that it brings any eigenspace $P_\lambda$ to an eigenspace $P_\epsilon$ with $\epsilon$ near zero. Take two  partial isometries $f_1$ and $f_2$ that, respectively, map the eigenspace $P_{\lambda_1}$ to $P_{\epsilon_1}$ and $P_{\lambda_2}$ to $P_{\epsilon_2}$ such that $\lambda_1\neq\lambda_2$ and $\epsilon_1\neq \epsilon_2$ then the operator $v_1+v_2$ is also partial isometry that maps $P_{\lambda_1}+P_{\lambda_2}$ to $P_{\epsilon_1}+P_{\epsilon_2}$ with $\epsilon$'s in a small neighborhood of zero. Adding such partial isometries and using a bijection from the eigenspace with $\lambda\in (-\Lambda,\Lambda)$ to the one with $\lambda\in (-\epsilon,\epsilon)$ we construct a partial isometry $v_\Lambda$ that compresses 
the spectrum of the relative modular operator from $(e^{-\Lambda},e^\Lambda)$ to a narrow interval near one $(e^{-\epsilon},e^\epsilon)$. 
Denote by 
$\Pi_\Lambda=\int_{-\epsilon}^\epsilon d\lambda P_\lambda$ the projection in the spectrum to $(e^{-\epsilon},e^\epsilon)$.
The state $P_0\ket{\Omega}$ is invariant under the action of the relative modular operator and the state $\Pi_\epsilon\ket{\Omega}$ for small $\epsilon$ is almost invariant in the sense that
\begin{eqnarray}
\left\|\lb \Delta_{\Psi|\Omega}^{1/2}-1\rb \Pi_\epsilon\ket{\Omega}\right\|\leq \lb e^{\epsilon/2}-1\rb\ .
\end{eqnarray}
At large $\Lambda$, we can approximate $\ket{\Omega}\simeq \Pi_\Lambda\ket{\Omega}$, therefore the state $v_\Lambda\ket{\Omega}$ is also almost invariant under the action of relative modular operator:
\begin{eqnarray}\label{ineq}
\|(\Delta_{\Psi|\Omega}^{1/2}-1)v_\Lambda \ket{\Omega}\|&&\leq \|(\Delta_{\Psi|\Omega}^{1/2}-1)v_\Lambda \Pi_\Lambda\ket{\Omega}\|\nn\\
&&+\|(1-\Pi_\Lambda)(\Delta_{\Psi|\Omega}^{1/2}-1)\ket{\Omega}\|\nn\\
&&\leq\lb e^{\epsilon/2}-1\rb +2 \|1-\Pi_\Lambda\|\ .
\end{eqnarray}
Intuitively, one expects that in the limit $\Lambda\to \infty$ the sequence of $v_\Lambda$ tends to a unitary operator as the right-hand-side of the inequality above becomes vanishingly small. We postpone a discussion of this limit to section \ref{connes} and conclude that in infinite dimensional systems, the states $\ket{\Omega}$ and $\ket{\Psi_\Omega}$ for which the modular operator has an entire continuous spectrum there exists a local unitary such that $\|\ket{\Psi_\Omega}-U\otimes U_J\ket{\Omega}\|\leq \epsilon$ for $\epsilon$ arbitrarily small.\footnote{Such states can be constructed in infinite dimensional systems with density matrices (so-called type I systems). However, not all states have the property above. We thank Roberto Longo for pointing this out to us.}  Since any purification of local state $\psi$ is related to the canonical one by a unitary in $\bar{\mA}$ we have 
\begin{eqnarray}
\inf_{U\in \mA, U'\in\bar{\mA}}\|UU'\ket{\Omega}-\ket{\Psi}\|=0
\end{eqnarray}
for $\ket{\Omega}$ and $\ket{\Psi}$ any two vectors in the Hilbert space.

The local algebra of QFT is the limit of a finite quantum system where for every state the individual probabilities $p_k$ go to zero and only the ratio $p_{k'}/p_k$ makes sense.\footnote{See \cite{witten2018aps} for the construction of the local algebra of quantum fields as the limit of finite quantum systems.} Every eigenvalue $p_{k'}/p_k$ repeats infinite times, therefore the sum over $\gamma$ in (\ref{modularop}) runs to infinity. In the continuum limit, it is more precise to replace the expression in these paranthesis in (\ref{specrel}) with a projection operator valued measure $P(d\lambda)$ and write
\begin{eqnarray}\label{specrel2}
\Delta_{\Psi|\Omega}=\int e^\lambda P(d\lambda)\ .
\end{eqnarray}
In \cite{connes1978homogeneity} Connes and Stormer used a generalization of the argument above to prove that for every pair of states $\ket{\Omega_U}$ and $\ket{\Psi}$ there exists a local unitary $U$ that satisfies $\|UU_J\ket{\Omega}-\ket{\Psi_\Omega}\|\leq \epsilon$ for $\epsilon$ arbitrarily small.
In other words, they showed that the strong distance of unitary orbits of any two local states is zero, i.e. $d_s([\omega],[\psi])=0$. We review the steps needed for the generalization in section \ref{connes}.  Unfortunately the argument in \cite{connes1978homogeneity} just proves the existence of a unitary $U$ that sews states. In the remainder of this section, we focus on the weak distance $d_w([\omega],[\psi])$ so that we can explicitly construct examples of unitaries that sew states of QFT. 

\section{Connes-Stormer Result}\label{connes}

In QFT, there are no local Hilbert spaces analogous to $\mathcal{H}_A$. Splitting a time-slice into two complementary regions $A$ and $\bar{A}$ there are local algebras on each region that we denote by $\mA$ and $\bar{\mA}$. They are isomorphic and together they generate the global algebra. In the problem of sewing local states we are only interested in pure global states that we refer to as vectors in the Hilbert space. The local state on $\mA$ is the restriction of the expectation values in the global vector $\ket{\Omega}$ to the local algebra $\omega(a)=\braket{\Omega|a\Omega}$ with $a\in\mA$. We can sew local states $\omega$ on $A$ to $\psi$ on $\bar{A}$ if there exists a global vector $\ket{\Phi}$ such that 
\begin{eqnarray}\label{sewing}
&&\forall a\in \mathcal{A}:\qquad\omega(a)=\braket{\Phi|a\Phi}\nn\\
&&\forall a'\in\mathcal{\bar{A}}: \qquad \psi(a')=\braket{\Phi|a'\Phi}\ .
\end{eqnarray}
We say an algebra 
$\mA=\mA_1\otimes \mA_2$ if for any local states $\omega$ on $\mA_1$ and $\psi$ on $\mA_2$ there exists a state $\phi$ on $\mA$ such that 
\begin{eqnarray}
&&\forall a\in \mathcal{A}_1:\qquad\omega(a)=\phi(a)\nn\\
&&\forall b\in\mathcal{A}_2: \qquad \psi(b)=\phi(b)\ .
\end{eqnarray}
 In QFT, as opposed to finite quantum systems, if $\bar{A}$ is the complement of $A$ the global algebra does not split into a tensor product $\mA\otimes\bar{\mA}$. 
The absence of tensor product between the algebra of region $A$ and that of $\bar{A}$ implies that it is impossible to sew all states perfectly. However, as we argue below it is possible to sew any two states with arbitrary precision.

Consider the set of vectors in the Hilbert space that reduce to $\omega$ on $\mA$. Analogous to the case of qudits, there exists a unique canonical global vector $\ket{\Omega}$ that is invariant under the anti-linear $J_\Omega$ that swaps $A$ and $\bar{A}$ \cite{haag2012local}. 
Given a second local state $\psi$ on $\mA$ there exists a unique global vector $\ket{\Psi_\Omega}$ that reduces to $\psi$ and is invariant under $J_\Omega$. We postpone a definition of $J_\Omega$ to the next section. In (\ref{Hilbertdis}), we expressed the Frobenius distance of local states on $\mA$ in terms of the Hilbert space distance of their corresponding global vectors. We take the Hilbert space distance as the definition of distance between local states
\begin{eqnarray}
d_F(\omega,\psi)=\|\ket{\Psi_\Omega}-\ket{\Omega}\|\ .
\end{eqnarray}
Rotating the local state by a local unitary corresponds to considering a new local state with $\omega_U(a)=\omega(U a U^\dagger)$. We are interested in the minimum distance in the unitary orbit distance of $\omega$:
\begin{eqnarray}\label{infqft}
\inf_{U\in\mA} d(\omega_U,\psi)=\inf_{U\in\mA}\|\ket{\Psi_\Omega}-\ket{U_\Omega}\|\ .
\end{eqnarray}
where $\ket{U_\Omega}=UU_J\ket{\Omega}$ is the canonical purification of $\omega_U$.
In the last section, we argued that if the modular operators of both $\ket{\Psi_\Omega}$ and $\ket{\Omega}$ have an entirely continuous spectrum the infimum above is zero.
We now generalize that argument to show that in QFT
the infimum above is zero for any pair of states. The key observation is that QFT is a special type of infinite quantum system for which 
the spectrum of the modular operator of any state is the entire $[0,\infty)$.\footnote{A quantum system with this property is called type III$_1$. The local algebra of QFT in any dimension is a type III$_1$ system.} 
If $U$ is the local unitary that makes the infimum in (\ref{infqft}) vanishingly small the state  $U\ket{\Omega}$ reduced to $\mA$ is arbitrarily close to $\psi$, whereas outside of $\mA$ the state is the same as $\omega$. 
The seeming contradiction with finite quantum systems has to do with the fact the there is no notion in QFT analogous to the eigenvalues of local states. There is only the modular operator whose spectrum is entirely continuous with no eigenvalues \footnote{We require an eigenvector to be normalizable.}. That is the reason underlying the divergences that appear in a naive calculation of unitary invariant measures of local states such as entanglement or Renyi entropies. 

An important step in the argument of the last section was the construction of the partial isometry $v_\Lambda$. There is no local Hilbert space $\mathcal{H}_A$ in QFT, therefore one needs to approximate the partial isometry $f_{kk'}=\ket{\alpha_{k}}\bra{\alpha_{k'}}$ that relates the projections $f_{kk}$ and $f_{k'k'}$.
The projection $f_{kk}$ is special in that it commutes with the density matrix $\psi$.
In QFT, the projections in the local algebra that commute with the modular operator play the role of $f_{kk}$, i.e. $[\Delta_\Psi,f_{kk}]=0$. The other special feature of $f_{kk}$ is that it is the smallest projection in the sense that there are no other non-zero projection $f'$ such that $f'\mathcal{H}\subset f_{kk}\mathcal{H}$. This aspect cannot be generalized to QFT.\footnote{A quantum system is called type I if and only if it has minimal projections in the sense defined here. QFT is not a type I system.}
To run the argument in QFT we start with two projections $f_k$ and $f_{k'}$ that commute with the modular operator $\Delta_\Psi$ and take the partial isometry that relates them as a replacement of $f_{kk'}$.\footnote{It is possible that there are no $f$ in the local algebra that commutes with the modular operator. We consider that case in section \ref{commutator}.}
We repeat the argument of the last section by considering two partial isometries of this type $v_1$ and $v_2$ and their corresponding projections, respectively, $f_i=v_i^\dagger v_i$ and $f'_i=v_iv_i^\dagger$ for $i=1,2$. If $f_1\perp f_2$ and $f'_1\perp f'_2$ the sum $v_1+v_2$ is also partial isometry with extended domain $(f_1+f_2)\mathcal{H}$ and range $(f'_1+f'_2)\mathcal{H}$. We  continue adding such partial isometries until either the domain or the range of the partial isometry becomes the whole Hilbert space, at which point the resulting operator is either an isometry $v^\dagger v=1$ or a co-isometry $vv^\dagger=1$.\footnote{Since the spectrum of $\Delta_\Psi$ is the uncountable set $[0,\infty)$ the rigorous argument is as follows: If we have two sets of projections $\tilde{f}-f>0$ and $\tilde{f'}-f'>0$ and partial isometries $v^\dagger v=f$ and $v v^\dagger=f'$ and $\tilde{v}^\dagger\tilde{v}=\tilde{f}$ and $\tilde{v}\tilde{v}^\dagger=\tilde{f'}$. we say $\tilde{v}\succ v$ if $\tilde{v} f=v$ and $\tilde{v}^\dagger f'=v^\dagger$. This define a partial order among partial isometries. Since every chain in this partially ordered set has a maximal element, by Zorn's lemma, there exists a maximal partial isometry that one proves to be either an isometry or a co-isometry \cite{connes1978homogeneity}.}
The isometry $v$ has the property that $\|\ket{\Psi_\Omega}-vv_J\ket{\Omega}\|\simeq 0$. The final step is to notice that in QFT any isometry is the limit of sequence of unitaries in the strong operator topology; see appendix \ref{AppC}. This finishes the proof that in QFT any state $\omega$ on $A$ can be sewn to any $\psi$ on $\bar{A}$.


\section{Modular theory and the cocycle}\label{modulartheory}
In this section, our goal is to go beyond systems with density matrices and define the appropriate generalizations of the relative modular operator, modular conjugation and the cocycle in a general quantum system.
The mathematical framework that accomplishes this is called the Tomita-Takesaki modular theory. See \cite{witten2018aps,lashkari2018modular} for a reveiew of the modular theory.
The remainder of this section applies to any general quantum system, from qubits to QFT.

Consider a local quantum system $A$ and the auxiliary quantum system $\bar{A}$ that purifies its state. Their corresponding algebras $\mA$ and $\bar{\mA}$ together generate a global algebra with a representation on a separable Hilbert space $\mathcal{H}$. The pure global states are vectors $\ket{\Omega}\in\mathcal{H}$. 
There are many vectors in the Hilbert space that reduce to $\omega$ on $\mA$. Acting on a vector with any isometry $W'\in\bar{\mA}$, i.e. $W'^\dagger W'=1_{\bar{A}}$ and $W'W'^\dagger=\pi\in\bar{\mA}$ a projection, leaves the local state invariant
\bea
\braket{\Omega|a\Omega}=\braket{W'\Omega|a W'\Omega}\ .
\eea
A vector $\ket{\Omega}$ has the Reeh-Schlieder property if the set of vector $a\ket{\Omega}$ is dense in the Hilbert space. 
Every local operator $a\in\mA$ has a hermitian conjugate operator $a^\dagger\in\mA$. We define the anti-linear operator $S_\Omega^A$ using its action on vectors  
\bea\label{tomita}
\forall a\in\mathcal{A}:\qquad S^A_\Omega a\ket{\Omega}=a^\dagger\ket{\Omega}\ .
\eea
For vectors with the Reeh-Schlieder property $S_\Omega^A$ in (\ref{tomita}) is densely defined. Hereafter, we refer to $S_\Omega^A$ as the closure of the operator above. The $S_\Omega^A$ is called the Tomita operator and is the starting point of modular theory.
The modular operator is the norm of the Tomita operator: $\Delta_\Omega=S_\Omega^\dagger S_\Omega$ where we have suppressed the region index $A$. The modular conjugation is the anti-linear operator $J_\Omega=\Delta_\Omega^{1/2}S_\Omega$. If $\ket{\Omega}$ has the Reeh-Schlieder property $J_\Omega$ is an anti-unitary.
The results of the modular theory that we need here are the following two facts:\footnote{See \cite{bratelli1982operator} for proofs of these statements.} 
\begin{enumerate}
    \item The operator $\Delta_\Omega^{it}$ generates a unitary flow in the algebra \begin{eqnarray}
    a\in\mathcal{A}\to a_\Omega(t)\equiv\Delta^{it}_\Omega a\Delta_\Omega^{-it}\in\mathcal{A}
    \end{eqnarray}
    \item The modular conjugation is an anti-linear map that sends every operator in the $\mA$ to $\bar{\mA}$: $a_J\equiv J_\Omega aJ_\Omega\in\bar{\mA}$. Every local state $\omega$ on $A$ has a unique canonical purification $\ket{\Psi_\Omega}$ satisfying: $J_\Omega\ket{\Psi_\Omega}=\ket{\Psi_\Omega}$.
\end{enumerate}  
In the case of a density matrix $\omega$ and its canonical purification $\ket{\Omega}$, the modular conjugation is the anti-linear swap operator we discussed earlier and the modular operator is $\Delta_\Omega=\omega\otimes \omega^{-1}$. The modular flow of operators in the algebra is given by $\omega^{it}a\omega^{-it}$.
The main difference in QFT is that, as opposed to the case of density matrices, the modular flow is not generated by unitaries in the algebra, but with the modular operator that is an operator in the global Hilbert space.

For two vectors $\ket{\Omega}$ and $\ket{\Psi}$ we define the relative Tomita operator and the relative modular operator as its norm
\begin{eqnarray}\label{relativeTomita}
    &&S_{\Psi|\Omega}a\ket{\Omega}=a^\dagger\ket{\Psi}, \qquad \Delta_{\Psi|\Omega}=S^\dagger_{\Psi|\Omega}S_{\Psi|\Omega}\ .
\end{eqnarray}
If we choose the canonical purification $\ket{\Psi_\Omega}$ then $\Delta_{\Psi_\Omega|\Omega}^{1/2}\ket{\Omega}=\ket{\Psi_\Omega}$. The definition of the weak and strong distance  in (\ref{weakstrong}) and definition of the cocycle $u_{\Psi|\Omega}(t)=\Delta_{\Psi|\Omega}^{it}\Delta_\Omega^{-it}$ generalize trivially.  However, we need to show that the cocycle is inside the algebra $\mA$.
To prove this, we use a trick introduced by Connes in \cite{connes1973classification}. 
We add a qubit to the local algebra $\mA\equiv \mA_A$ so that the new algebra $\mA_{AQ}$ is generated by $a\otimes \ket{i}\bra{j}$ with $\ket{i}$ a vector in the Hilbert space of the qubit $Q$. We also add a qubit $\bar{Q}$ to the algebra of the complementary region $\bar{\mA}\equiv \mA_{\bar{A}}$ so that we can purify local states of $AQ$ in $AQ\bar{A}\bar{Q}$.
Consider the global state 
\bea
\ket{\Theta}_{AQ\bar{A}\bar{Q}}=\ket{\Omega}_{A\bar{A}}\otimes \ket{00}_{Q\bar{Q}}+\ket{\Psi}_{A\bar{A}}\otimes \ket{11}_{Q\bar{Q}}
\eea
where $\ket{\Omega}$ and $\ket{\Psi}$ are global field theory states which we assume to be Reeh-Schlieder.\footnote{It is straightforward to relax this assumption. In QFT, the set of states with the Reeh-Schlieder property is dense in the Hilbert space and includes the vacuum \cite{witten2018aps,haag2012local}.} 
The combined local state of $AQ$ is $\theta_{AQ}=\omega_A\otimes \ket{0}\bra{0}_Q+\psi_A\otimes\ket{1}\bra{1}_A$.
The Tomita operator for this state acts as $S_\Theta x
 \ket{\Theta}=x^\dagger \ket{\Theta}$ with $x\in \mA_{AQ}$. 
The Tomita operator and the modular conjugation of this state decompose according to 
\bea
S_\Theta=\begin{pmatrix}
  S_\Omega & & & \\
&  0 & S_{\Psi|\Omega}  &\\ 
&  S_{\Omega|\Psi} &0 &\\
& & & S_{\Psi}\\
 \end{pmatrix}
\eea
and
\bea
\Delta_\Theta=\begin{pmatrix}
  \Delta_{\Omega} & & & \\
&  \Delta_{\Omega|\Psi} & &\\ 
&  & \Delta_{\Psi|\Omega}  &\\
& & & \Delta_\Psi\\
 \end{pmatrix}\ .
\eea
The cocycle $u_{\Psi|\Omega}(t)$ is simply the modular flow of the operator $1_A\otimes \ket{0}\bra{1}_Q$ in state $\ket{\Theta}$:
\bea
\Delta_\Theta^{it}(1\otimes \ket{0}\bra{1}_Q\otimes 1_{\bar{A}\bar{Q}})\Delta_\Theta^{-it}&&=\Delta_\Omega^{it}\Delta_{\Psi|\Omega}^{-it}\otimes \ket{00}\bra{10}\nn\\
&&+\Delta_{\Omega|\Psi}^{it}\Delta_\Psi^{-it}\otimes \ket{01}\bra{11}\nn\ .
\eea
Since the modular flow has to remain inside $\mathcal{A}_{AQ}$ we find that the cocycle belongs to the local algebra of field theory in $A$ and satisfies
$u_{\Psi|\Omega}(t)= \Delta_{\Psi|\Omega}^{it}\Delta_\Omega^{-it}=\Delta_\Psi^{it}\Delta_{\Omega|\Psi}^{-it}$.\footnote{If we start with a three-level system for $Q$ instead of a qubit the same argument establishes the so-called cocycle chain rule: $u_{\Psi|\Omega}(t)u_{\Omega|\Phi}(t)=u_{\Psi|\Phi}(t)$ \cite{connes1973classification} which in the case $\Phi=\Psi$ gives $\Delta_{\Psi|\Omega}^{it}\Delta_\Omega^{-it}=\Delta_\Psi^{it}\Delta_{\Omega|\Psi}^{-it}$.} 
More generally, the modular flow of the operator $V\otimes \ket{0}\bra{1}$ is
\begin{eqnarray}\label{relmodflow}
&&\Delta_\Theta^{it}\lb V\otimes \ket{0}\bra{1}_Q\rb\Delta_\Theta^{-it}=V_\Omega(t)u_{\Psi|\Omega}(t)^\dagger\otimes \ket{0}\bra{1}\nn,
\end{eqnarray}
which is the conjugate of the operator we found in section \ref{secinv} to sew states in the large $t$ limit.
If both states are Reeh-Schlieder the cocycle generates a unitary flow in the algebra.
For an operator in the complementary region $a'\in \mA_{\bar{A}}$ we have 
$u^\dagger_{\Psi|\Omega}a'u_{\Psi|\Omega}=a'$
which further implies 
\bea\label{relflow}
&&\Delta_{\Psi|\Omega}^{it} a'\Delta_{\Psi|\Omega}^{-it} =\Delta_\Omega^{it} a'\Delta_\Omega^{-it}\nn\\
&&\Delta_{\Psi|\Omega}^{it} a \Delta_{\Psi|\Omega}^{-it} =\Delta_\Psi^{it} a\Delta_\Psi^{-it}
\eea
where in the second line we have used the relation $\Delta^{\bar{A}}_{\Omega|\Psi}=(\Delta^A)_{\Psi|\Omega}^{-1}$ \cite{lashkari2018modular}.

As an example, consider state $u_{\Psi|\Omega}(t)\ket{\Omega}$ in the large $t$ limit. From (\ref{relmodflow}) we have
\begin{eqnarray}
\lim_{t\to\infty}\braket{\Omega|u(t)^\dagger a' u(t)|\Omega}&&=\braket{\Omega|a\Omega}\nn\\
\lim_{t\to\infty}\braket{\Omega|u(t)^\dagger a u(t)|\Omega}
&&=\lim_{t\to \infty}\omega_{u(t)}(a)\ .
\end{eqnarray}
In section \ref{secinv} we found that the weak distance $\lim_{t\to \infty}d_w(\omega_{u(t)},\psi)=0$. That is the sense in which the state is the same as $\psi$ on $A$. 

Consider the algebra of half-space ($x>|t|$) in the vacuum state of QFT in any dimension. The modular operator is $e^{-2\pi K_x}$ where $K_x$ is the boost operator in the $x$ direction \cite{bisognano1976duality}. 
If we pick $\ket{\Psi}$ to be the local state of the vacuum QFT on half space the modular flow $\Delta_{\Psi}^{it}$ is the boost operation. The authors of \cite{balakrishnan2019general} argued that $\Delta_\Psi^{it}\ket{\chi}$ for some $\ket{\chi}$ is an infinitely boosted state that should look like vacuum. Here, we presented a generalization of that argument to an arbitrary state $\ket{\Psi}$.

A benefit of enlarging the local algebra by a qubit is that it allows us to write the Frobenius distance measure in equation (\ref{disFro}) in terms of the norm of a commutator between the local density matrix and an observable. In the next section, we use this rewriting to study invariant states of relative modular operator.

\section{Commutators and invariant states}\label{commutator}

In section \ref{connes}, the projections that commute with the modular operator and the invariant states of relative modular operator played an important role in our construction of the unitary that sew states. Here, we discuss such projections in more detail. Let us go back to the Hilbert space of system $AQ$ and consider the density matrix $\theta_{AQ}=\omega_A\otimes\ket{0}\bra{0}_Q+\psi_A\otimes\ket{1}\bra{1}_Q$. For any $a\in \mA_A$ we can write
\begin{eqnarray}
&&(\sqrt{\omega}a-a\sqrt{\psi})\otimes \ket{0}\bra{1}_Q=\left[\sqrt{\theta_{AQ}}, a\otimes \ket{0}\bra{1}_Q\right]\ .
\end{eqnarray}
We rewrite the Frobenius distance of density matrices as
\begin{eqnarray}
d_F(\omega,\psi)^2&&=\frac{1}{2}\left\|\left[\sqrt{\theta_{AQ}},1_A\otimes \sigma^X_Q\right]\right\|_F^2\nn\\
&&=-\frac{1}{2}\text{tr}\lb \left[\sqrt{\theta_{AQ}},1_A\otimes  \sigma^X_Q\right]^2\right),
\end{eqnarray}
where $\sigma^X_Q=\ket{0}\bra{1}+\ket{1}\bra{0}$ is the $X$ Pauli matrix in the algebra of the qubit $Q$. More generally, for an arbitrary operator $a\in\mA$ we have
\begin{eqnarray}\label{skew1}
&&\frac{1}{2}\|\sqrt{\omega}a-a\sqrt{\psi}\|_F^2=\|a\ket{\Psi_\Omega}-a^\dagger_J\ket{\Omega}\|\nn\\
&&=-\frac{1}{2}\text{tr}\lb \left[\sqrt{\theta},x_a\right]^2\right)=\left\|\lb \Delta_\Theta^{1/2}-1\rb x_a\ket{\Theta}\right\|,
\end{eqnarray}
where $x_a\equiv a\otimes \ket{0}\bra{1}+a^\dagger\otimes \ket{1}\bra{0}$ is in the algebra of $AQ$.
Our distance measure is the Frobenius norm of the commutator of density matrix $\theta$ and an observable $x_a$, otherwise known as the Wigner-Yanase skew information \cite{wigner1963z}. For unitary orbits we rewrite the infimum distance as
\begin{equation}
    d_F([\omega],[\psi])^2=-\frac{1}{2}\inf_U\text{tr}\lb\left[\sqrt{\theta},x_U\right]^2\rb,
\end{equation}
where $x_U=U\otimes \ket{0}\bra{1}+U^\dagger \ket{1}\bra{0}$ is a self-adjoint unitary operators. 
If there exists an $x_U$ that commutes with $\theta$ the density matrices $\omega$ and $\psi$ are simultaneously diagonalizable. If $x_U$ almost commutes with $\theta_{AQ}$ then $U\omega U^\dagger$ becomes close to $\psi$. The almost invariant states of relative modular operator are prepared by $x_U$ that almost commute with $\theta$.

Consider the density matrix $\omega=\sum_k e^{-\lambda_k} \ket{k}\bra{k}$ and its canonical purification $\ket{\Omega}$. The orthonormal projection $e_k=\ket{k}\bra{k}$ commute with $\omega$. If there are degenerate eigenvalues, i.e. $\lambda_k=\lambda_{k'}$, then the partial isometry $\ket{k}\bra{k'}$ also commutes with $\omega$. The sub-algebra of all operators that commute with $\omega$ is called the centralizer of $\omega$.
If the algebra has a center $Z$, the center is in the centralizer of all density matrices. Using the cyclicity of trace we find that any operator $h$ that commutes with $\omega$ satisfies: $\text{tr}([\omega,h]a)=\text{tr}(\omega[h,a])=0$ for all $a\in\mA$. In QFT, we define the centralizer as the set of $h$ such that $\omega([h,a])=0$ for all $a\in\mA$. 
Any operator $h$ in the centralizer of $\omega$ commutes with the modular operator, and is invariant under modular flow \cite{pedersen1973radon}:
\begin{eqnarray}
&&(\Delta^{1/2}_\Omega-1)h\ket{\Omega}=0\nn\\
&&\Delta_\Omega^{it} h \Delta_\Omega^{-it}=h\ .
\end{eqnarray} 
Acting with $h$ in the centralizer on $\ket{\Omega}$ creates a new invariant state of $\Delta_\Omega$. Moreover, every invariant vector $\ket{h}$ is $h\ket{\Omega}$ for some $h$ affiliated with the centralizer \cite{pedersen1973radon}.\footnote{Affiliated with $\mA$ means that it commutes with $\bar{\mA}$. If $\bar{\mA}$ has trivial center any bounded operator that is affiliated with $\mA$ is in $\mA$.} Therefore, to understand invariant states of modular operator it suffices to study the centralizer of the state. 

So far we have only discussed the invariant states of the relative modular operator $\Delta_{\Psi|\Omega}$ that are locally prepared as $u\ket{\Omega}$. It is natural to ask what does a general invariant state of the relative modular operator look like.
Consider the state $UU'\ket{\Omega}$ with $U\in\mA$ and $U'\in\bar{\mA}$. Its relative modular operator is $\Delta_{\Omega|UU'\Omega}=U'\Delta_\Omega (U')^\dagger$ \cite{lashkari2019constraining}, therefore $U'\ket{\Omega}$ is its invariant state. It is a state that looks like $\ket{\Omega}$ on $A$ and $\ket{UU'\Omega}$ on $\bar{A}$. As we showed in this work, every state $\ket{\Psi}$ is well-approximated by some $UU'\ket{\Omega}$. Hence, it is tempting to think that all states $\ket{\Omega_\Psi}$ that are invariant under $\Delta_{\Omega|\Psi}$ look like $\ket{\Omega}$ on $A$ and $\ket{\Psi}$ on $\bar{A}$, and only differ by their long-range correlations across $A\bar{A}$. We show below that this is incorrect if the centralizers of the states are non-trivial.

Every operator $h$ that commutes with $\Delta_\Omega$ also commutes with $\Delta_{\Omega|\Psi}$:
\bea
\Delta_{\Omega|\Psi}^{it}h\Delta_{\Omega|\Psi}^{-it}=\Delta_\Omega^{it} h\Delta_{\Omega}^{-it}=h
\eea
where we used (\ref{relflow}). If $\ket{\Omega_\Psi}$ is an invariant state of relative modular flow so is $h\ket{\Omega_\Psi}$. This implies that if $UU_J\ket{\Omega}\simeq \ket{\Psi_\Omega}$ then $(h U)(h U)_J\ket{\Omega}\simeq hh_J\ket{\Psi_\Omega}$.
The physical interpretation of an arbitrary invariant state $\ket{\Omega_\Psi}$ is that it is a state that is the same as $h\ket{\Omega}$ with respect to $\mA$ and $\tilde{h}\ket{\Psi}$ with respect to $\bar{\mA}$ where $h$ is in the centralizer of $\ket{\Omega}$ and $\tilde{h}$ is in the centralizer of $\ket{\Psi}$. In the state $\ket{\Omega_\Psi}$ the flow $\Delta_\Omega^{it}$ generates a symmetry of $\mA$ and $\Delta_\Psi^{it}$ generates a symmetry of $\bar{\mA}$:
\bea
&&\Delta_{\Omega|\Psi}\ket{\Omega_\Psi}=\ket{\Omega_\Psi}\nn\\
&&\braket{\Omega_\Psi|a\Omega_\Psi}=\braket{\Omega_\Psi| \Delta_{\Omega|\Psi}^{it} a \Delta_{\Omega|\Psi}^{-it}\Omega_\Psi}=\braket{\Omega_\Psi|\Delta_\Omega^{it} a\Delta_{\Omega}^{-it}\Omega_\Psi}\nn\\
&&\braket{\Omega_\Psi|a'\Omega_\Psi}=\braket{\Omega_\Psi| \Delta_{\Omega|\Psi}^{it} a' \Delta_{\Omega|\Psi}^{-it}\Omega_\Psi}=\braket{\Omega_\Psi|\Delta_\Psi^{it} a\Delta_\Psi^{-it}\Omega_\Psi}\nn
\eea
for all $a\in\mathcal{A}$ and $a'\in\mathcal{\bar{A}}$. From the point of view of $\mA$ the state $\ket{\Omega_\Psi}$ is $h\ket{\Omega}$ and from the point of view of the algebra  $\bar{\mA}$ the state is $\tilde{h}\ket{\Psi}$. 

In the vacuum of QFT restricted to half-space the modular operator is the boost.
Vacuum is the only boost invariant state, therefore its centralizer is trivial. The modular flow in such states is ergodic in the sense that in the large time limit $\Delta_\Omega^{it}\ket{\chi}$ converges to vacuum in the weak norm for any state $\ket{\chi}$.
See \cite{longo1982algebraic} for a review.
The algebra of QFT is not the tensor product of $\mathcal{A}\otimes\mathcal{\bar{A}}$ which means that  not all $\ket{\Omega}$ and $\ket{\Psi}$ can have exact invariant states of their relative modular operator $\ket{\Omega_\Psi}$. However, there are always states that are almost invariant with arbitrary precision. We take this as evidence that, in general, the exact state $\ket{\Omega_\Psi}$ that sews $\omega$ on the left with $\psi$ on the right has a singularity (shockwave) at the boundary where the states are sewn together.
As we saw this singular state $\ket{\Omega_\Psi}$ can be approximated arbitrarily well with normalizable states by acting with unitaries from the algebra. The approximation states are invariant under $\Delta_\Psi^{it}$ in $\bar{\mA}$ and almost invariant under $\Delta_\Omega^{it}$ in $\mA$. 

The centralizer of the vacuum of QFT is trivial. To apply the argument of section \ref{secinv} we need to find projections that are approximately in the centralizer of the vacuum. In this case, we need to find approximately boost-invariant projections in the algebra of the half-space.
Once again, we start with finite quantum systems to obtain intuition. 
It is straightforward to construct operators that commute with density matrix $\omega$ using the modular flow $a_\omega(t)=\omega^{it}a\omega^{-it}$. The idea is to integrate $a_\omega(t)$ over  $t$ to kill the off-diagonal terms:
\begin{eqnarray}
\left[\sqrt{\omega},\int_{-\infty}^\infty dt \: a_\omega(t)\right]&&=\sum_{k,k'}\int dt\:  e^{it(\lambda_k-\lambda_{k'})}a_{kk'}\:\left[\sqrt{\omega},\ket{k}\bra{k'}\right]\nn\\
&&=\sum_k a_{kk}\left[\sqrt{\omega},\ket{k}\bra{k}\right]=0\ .
\end{eqnarray}
The operator $\int_{-\infty}^\infty dt\: a_\omega(t)$ is the zero Fourier mode of $a_\omega(t)$. A general Fourier mode $\hat{a}_\omega(l)=\int_{-\infty}^\infty dt \:e^{i t l} a_\omega(t)$ satisfies 
\begin{eqnarray}
    \left[\sqrt{\omega},\hat{a}_\omega(l)\right]&&=\sum_{k} a_{k(k-l)}\:\left[\sqrt{\omega},\ket{k}\bra{k-l}\right]\nn\\
   && =\sum_k a_{k(k-l)} e^{-\lambda_k/2}\lb 1-e^{l/2} \rb \ket{k}\bra{k_l}\nn,
\end{eqnarray}
where $\ket{k_l}$ is the eigenvector with $\lambda_{k_l}=\lambda_k-l$. If there are no such eigenvector $\hat{a}_\omega(l)=0$. In a finite quantum system the set of non-zero frequency modes is discrete, and the smallest frequency $l$ corresponds to is the smallest eigenvalue gap. If the spectrum of the modular operator we have chosen is the full positive line all Fourier frequencies are non-zero and one can choose the Fourier mode $\hat{a}_\omega(\epsilon)$ for some $\epsilon\ll 1$. Such a mode almost commutes with $\omega$:
\begin{eqnarray}
\|[\sqrt{\omega},\hat{a}_\omega(\epsilon)]\|_F\leq  \epsilon\|a\|\|\sqrt{\omega}\ket{k}\bra{k_\epsilon}\|_F\leq  \epsilon\|a\|\ .
\end{eqnarray}
If we take a set of Fourier modes $|l|<\epsilon$ and $g(l)$ independent of $\epsilon$ the operator $\int_{|l|\leq \epsilon} dl g(l) \hat{a}_\omega(l)$ almost commutes with $\omega$.
Fourier transforming back we find that for any square-integrable function $g(t)$ whose Fourier transform $g(l)$ is restricted to $|l|\leq \epsilon$ the operator $g_\omega(a)=\int dt \: g(t) a_\omega(t)$ almost commutes with $\sqrt{\omega}$. It is self-adjoint if $g(t)=g(-t)$. 
The Fourier modes of modular flow have been discussed in connection with the holographic duality in \cite{faulkner2017bulk}. 


\section{Sewing multiple states}\label{sewmulti}

Finally, we come to the case of $n$ subsystems.
Consider a collection of $n$ density matrices $\{\omega^{(1)},\cdots, \omega^{(n)}\}$ each corresponding to a qubit. The qubits can be sewn together if there exists a global pure state $\ket{\Omega}$ such that $\omega^{(i)}=tr_{i\neq j}\ket{\Omega}\bra{\Omega}$ for all $i$. Local unitaries that rotate each qubit $U_i \omega^{(i)}U_i^\dagger$ have no effect on whether or not the density matrices can be sewn. The constraints on sewing $\omega^{(i)}$ only depend on their eigenvalues. Expand each density matrix in its diagonal eigenbasis:
\begin{eqnarray}
\omega^{(i)}=\sum_i \lambda_i \ket{0}_i\bra{0}_i+(1-\lambda_i)\ket{1}_i\bra{1}_i
\end{eqnarray}
with $0\leq \lambda_i\leq 1/2$. The obstruction to sew density matrices is a constraint on the set of eigenvalues $\lambda_i$. We briefly review the derivation of these constraints in \cite{bravyi2003requirements}.

In the case of two qubits, one can sew the qubits if and only if they have the same spectrum, i.e. $\lambda_1=\lambda_2$.
Next, consider the case of three qubits. The projection $e_i=\ket{0}\bra{0}_i$ satisfy the identity: $e_1+e_2=1_{12}+e_1e_2-(1-e_1)(1-e_2)$.
This identity implies that if $\ket{\Omega}$ is a global vectors that sews them together 
\begin{eqnarray}\label{eigen}
\lambda_1+\lambda_2=\braket{\Omega|e_1+e_2|\Omega}\geq 1-\text{tr}\lb \omega^{(12)}(1-e_1)(1-e_2)\rb,\nn
\end{eqnarray}
where we have discarded the positive term $\braket{\Omega| e_1e_2\Omega}$. Since $\ket{\Omega}$ is pure the eigenvalues of $\omega^{(12)}$ are the same as those of $\omega^{(3)}$. Its largest eigenvalue is $1-\lambda_3$ which means $\braket{\Phi|\omega^{(12)}\Phi}\leq 1-\lambda_3$ for all $\ket{\Phi}$ in the Hilbert space of three qubits. The projector $(1-e_1)(1-e_2)$ is rank one, therefore $\text{tr}\lb \omega^{(12)}(1-e_1)(1-e_2)\rb\leq 1-\lambda_3$. Plugging this back into (\ref{eigen}) we find 
\begin{eqnarray}
\lambda_1+\lambda_2\geq \lambda_3\ .
\end{eqnarray}
Permuting the qubits and repeating the same argument  gives two more constraints:
\begin{eqnarray}\label{cons}
\sum_{i\neq j}\lambda_j\geq \lambda_i\ .
\end{eqnarray}
Any three density matrices with $\{\lambda_1,\lambda_2,\lambda_3\}$ that satisfy the above inequality can be sewn together in a global vector that is non-unique. 

This argument generalizes to $n$-qubits in a straightforward manner, and the final constraint on the eigenvalues $\lambda_i$ are the same as (\ref{cons}) but with $i=1,\cdots n$. The eigenvalues of local density matrices of an $n$-partite system that can be sewn together form a convex polytope.
See \cite{walter2014multipartite} for a review of the generalization to arbitrary finite dimensional systems.

We now turn to quantum field theory.
Starting with any global state $\ket{\Psi}$ acting with local unitaries in region $A_1$ we can prepare an arbitrary local state $\omega_1$ on $A_1$. Then, we act locally on $A_2$ and prepare $\omega_2$, and repeat this for all subregions to $A_n$ to obtain a global state that sews local states $\omega_i$ for $i=1,\cdots, n$. There are no constraints on sewing $\omega_i$.
For instance, using local unitaries we can make every state look like vacuum on each $A_i$. Every state  $\ket{\Psi}$ has the same entanglement structure as $U_1U_2\cdots U_n\ket{\Psi}$. By classifying all states that sew vacuum reduced states $\omega_i$ together we learn about various forms of multi-partite entanglement than can appear in QFT. 

Consider the case $n=3$ in QFT with the region $A_2$ separating $A_1$ and $A_3$. We take the mutual information between $A_1$ and $A_3$ in the vacuum as a measure of entanglement between $A_1$ and $A_3$. This mutual information is the same as the relative entropy of $\omega_{13}$ with respect to $\omega_1$ and $\omega_3$:
\begin{eqnarray}
I(A_1:A_3)=S(\omega_{13}\|\omega_1\otimes\omega_3)\ .
\end{eqnarray}
Since the relative entropy on $A_1A_3$ does not depend on the purification we write
\begin{eqnarray}
S_{vac}(\omega_{13}\|\omega_1\otimes\omega_3)=-\braket{\Omega|\log\Delta_{\omega_1\otimes\omega_3|\omega_{13}}|\Omega}\ .
\end{eqnarray}
Any other global state that sews $\omega_i$s has the form $\ket{\tilde{\Omega}}\simeq U_2U_{13}\ket{\Omega}$. The mutual information in this state is
\begin{equation}
    S(\tilde{\omega}_{13}\|\omega_1\otimes\omega_3)=-\braket{\Omega|U_{13}^\dagger\log\Delta_{\omega_1\otimes\omega_3|\omega_{13}}U_{13}|\Omega}\ .
\end{equation}
The vacuum relative modular operator $\Delta_{\omega_1\otimes \omega_3|\omega_{13}}$ teaches us about the mutual information in a dense set of states $\ket{\tilde{\Omega}}$. The same principle extends to tri-partite entanglement and other measures of multi-partite entanglement for larger $n$ that are invariant under local unitaries. 

It is tempting to conclude from the discussion above that the vacuum relative modular operators $\Delta_{\omega_1\otimes\cdots\omega_n|\omega_{1\cdots n}}$ contain the information about the multi-partite entanglement of all states. This is incorrect because the relative modular operator and its logarithm are unbounded and if we have a sequence of states $\psi_n\to \psi$ the relative modular operators $\lim_n\Delta_{\psi_n|\psi}$ need not converge. In fact, the relative entropy is not continuous, but just lower semi-continuous \cite{araki1976relative}:
\begin{eqnarray}
S(\phi\|\psi)\leq \lim_nS(\phi\|\psi_n)\ .
\end{eqnarray}
We postpone the study of the implications of sewing for the theory of multi-partite entanglement in QFT to future work.



\section{Discussion}
 In this work, we showed that in quantum field theory any collection of local reduced states in non-overlapping regions can be sewn together with arbitrary precision. We argued that the local unitary that acts on $\ket{\Psi}$ and creates the vacuum state $\omega$ is the cocycle $u_{\Omega|\Psi}(t)$ in the large $t$ limit.
Ideas similar to sewing states of quantum field theory have appeared in various context, recently \cite{VanRaamsdonk:2018zws,Marolf:2019zoo}. It would be interesting to explore the connection between our algebraic sewing prescription and sewing Euclidean path-integrals discussed in the literature.
Here, we addressed the problem of sewing one-body local states of in a global pure state. The collection of one-body density matrices can be understood as a mean-field approximation to the state \cite{bravyi2003requirements}. The problem can be generalized to sewing multi-body local states with overlapping regions, called the quantum marginal problem. See \cite{walter2014multipartite} for a discussion of the quantum marginal problem in the general setup. It is interesting to study the generalized marginal problem in QFT \cite{lashkari2019entanglement}. It is worthwhile to note that the sewing argument presented in this paper applied generically to any two states of a type III$_1$ von Neumann algebra, independent of which quantum field theory they belong to. It would be interesting to study the physics of sewing states of different QFT and its connection with boundary states in a QFT.

In a QFT with a global symmetry the algebra of charge-neutral operators can be enlarged by the generator of the symmetry group on $A$ \cite{Furuya:2020wxf}. The modular operator for the charge-neutral subalgebra has non-trivial centralizers that correspond to $\oplus \lambda_r 1_r$ where $1_r$ is the identity operator in the irreducible representation $r$ of the symmetry group. From our work here, it follows that if we take the infinite time limit of the cocycle $u_{\Psi|\Omega}(t)$ computed with the relative entropies defined with respect to the invariant algebra any excitation can be washed away (the state $\psi$ becomes locally like the vacuum $\omega$) except for centralizers that correspond to the charges. It is tempting to speculate that this procedure can be used to formally define a theory of hydrodynamics where all local physics is averaged out except for the conserved charges. 




\section{Acknowledgements}
We are greatly indebted to Edward Witten who first pointed out the Connes-Stormer theorem to us.
We would also like to thank Martin Argerami, Thomas Faulkner, Adam Levine, Roberto Longo, Raghu Mahajan, Thomas Sinclair, Mark Van Raamsdonk and Feng Xu for valuable discussions.

\bibliographystyle{JHEP}

\bibliography{localUnit}

\appendix

\section{Distance measures and unitary orbits}\label{AppA}
 
There are various distance measures one can introduce on the space of density matrices $\rho$ and $\omega$.  A few well-known distance measures commonly used in information theory are:
\bea\label{distance}
&&d_T(\rho,\omega)^2=\frac{1}{2}\|\rho-\omega\|^2\nn\\
&&d_B(\rho,\omega)^2=1-\|\sqrt{\rho}\sqrt{\omega}\|\nn\\
&&d_F(\rho,\omega)^2=\frac{1}{2}\|\sqrt{\omega}-\sqrt{\rho}\|_F^2\nn\\
&&d_R(\rho,\omega)^2=1-e^{-\frac{1}{2}S(\rho\|\omega)}\nn\\
&&S(\rho\|\omega)=\text{tr}(\rho \log \rho)-tr(\omega \log\rho)
\eea
where $\|X\|=\text{tr}(|X|)$.
In this work, we primarily used the distance $d_F(\rho,\omega)$. 
The trace distance $d_T$, the Bures distance $d_B$ and the distance $d_F$ are symmetric in their arguments. Bures distance is a metric, and is continuous with respect to the trace distance.\footnote{That is to say it can be bounded from above by trace distance.} It is closely related to quantum fidelity $F(\omega,\rho)=\|\sqrt{\omega}\sqrt{\rho}\|$.  All four distance measures vanish if and only if $\rho=\omega$, and remain invariant under simultaneous rotation of both states, i.e. $d(\rho,\omega)=d(U\rho U^\dagger, U\omega U^\dagger)$. Their relationship can be summarized with the inequality:
\bea\label{distanceineq}
0\leq d_T\leq d_B\leq d_F\leq d_R\leq 1
\eea
The first inequality is the Fuchs–van de Graaf inequality \cite{nielsen2002quantum}. The second inequality is due to the fact that $\text{tr}(\sqrt{\rho}\sqrt{\omega})\leq \|\sqrt{\rho}\sqrt{\omega}\|$. The last inequality uses $S(\rho\|\omega)\geq -2\ln\text{tr}(\sqrt{\rho}\sqrt{\omega})$ \cite{carlen2014remainder}.

\section{Skew information and relative entropy}\label{AppB}

In defining the skew information in (\ref{skew1}) we can use $p$-norms. This generalization of the Wigner-Yanase information is due to Dyson. For a self-adjoint operator $x$ and $\omega$ a density matrix purified canonically in $\ket{\Omega}$ the Wigner-Yanase-Dyson $p$-skew information is
\begin{eqnarray}
I_p(\omega, x)=-\text{tr}\lb [\sigma^p,x][\sigma^{1-p},x]\rb\ .
\end{eqnarray}
In terms of the canonical purification $\ket{\Omega}$ of the density matrix the skew information is
\bea
I_p(x,\Omega)=\braket{\Omega|x \lb \Delta_\Omega^p+\Delta_\Omega^{1-p}-1-\Delta_\Omega\rb x |\Omega}\ .
\eea
The derivative at $p=0$ is
\bea
\lim_{p\to 0}I_p(x,\Omega)=2\braket{\Omega|x\log\Delta_\Omega x|\Omega}\ .
\eea
If $x$ is a unitary operator then this is the relative entropy
\begin{eqnarray}
S(u^\dagger\Omega|\Omega)=-\braket{\Omega|\log\Delta_{u^\dagger\Omega|\Omega}|\Omega}\ .
\end{eqnarray}
The $p$-skew information is symmetric under $p\to 1-p$ with the symmetric point corresponding to the Wigner-
Yanase measure of section \ref{commutator}.
 


\section{Isometries as a limit of unitaries}\label{AppC}


In this section, we argue that in QFT the unitary orbit of any state passes through subspaces $\pi\mathcal{H}$ for any projection $\pi\in\mathcal{A}$. There are two parts to the argument. 
First, we observe that in QFT (in general, any type III von Neumann algebra) for every projection $\pi$ there exists an isometry $w$ such that $w w^\dagger=\pi$ and $w^\dagger w=1$ \cite{haag2012local}. The state $\ket{\Phi_w}=w\ket{\Phi}$ is an eigenstate of $\pi$ for any $\ket{\Phi}$. Therefore, by acting with an isometry we can bring any state to the subspace $\pi\mathcal{H}$. Second, we notice that in QFT (any type III algebra) any isometry $w$ is a limit of unitary operators $u_n$ in strong operator norm, i.e. $\lim_n\|w-u_n\|=0$.\footnote{We learned the argument presented here from Martin Argerami.} To see this, consider a sequence of projections $\pi_n$ in the algebra that converge to identity: $\lim_n \|1-\pi_n\|=0$. For any $n$ the two projections $1-\pi_n$ and $1-w \pi_n w^\dagger$ belong to the algebra. A type III algebra is one in which for any two projections $\pi$ and $\tilde{\pi}$ there exists an operator $v$ such that $v^\dagger v=\pi$ and $vv^\dagger=\tilde{\pi}$. Therefore, there exists a partial isometry $v_n$ such that $v_n^\dagger v_n=1-\pi_n$ and $v_n v_n^\dagger=1-w \pi_n w^\dagger$. The operator $\pi_n w^\dagger v_n=0$ is zero because its norm vanishes: $(\pi_n w^\dagger v_n)(v_n^\dagger w \pi_n)=0$. Similarly, we have $v_n\pi_n=0$. If we define the operator $u_n=w \pi_n+v_n$ we find that it is a unitary operator for all $n$:
\bea
&&u_n^\dagger u_n=1+\pi_n w^\dagger v_n+v_n^\dagger w\pi_n=1\nn\\
&&u_n u_n^\dagger=1+w\pi_n v_n^\dagger+v_n \pi_n w^\dagger=1\ .
\eea
The sequence of unitaries $u_n$ tends to $w$ in strong operator topology, i.e. $\lim_n \|w-u_n\|=0$.

We have established that there exists a sequence of unitaries that  bring any state to the eigenspace of any projector $\pi$ in the algebra: $ \pi \lb \lim_n u_n\ket{\Phi}\rb=\lim_n u_n \ket{\Phi}$. Since this can be done for any two states, one might wonder if any two states can be brought arbitrarily close with unitaries. This happens only in a type III$_1$ algebra.







\end{document}